# INVESTIGATION OF OSL SIGNAL OF RESISTORS FROM MOBILE PHONES FOR ACCIDENTAL DOSIMETRY

A. Mrozik, B. Marczewska, P. Bilski, W. Gieszczyk

*Institute of Nuclear Physics PAN, Radzikowskiego 152, 31-342 Krakow, Poland*

## HIGHLIGHTS

Impact of a mobile phone mode (switched on/off) on absorbed dose by resistors was showed.

The influence of the temperature during irradiation on absorbed dose was measured.

Dose distribution inside of a mobile phone was performed.

Fading factor of resistors was calculated.

## ABSTRACT

Resistors from mobile phones, usually located near the human body, are considered as individual dosimeters of ionizing radiation in emergency situations. The resistors contain $Al_2O_3$, which is optically stimulated luminescence (OSL) material sensitive to ionizing radiation. This work is focused on determination of dose homogeneity within a mobile phones which was carried out by OSL measurements of resistors placed in different parts inside the mobile phone. Separate, commercially available resistors, similar in the shape and size to the resistors from circuit board of the studied mobile phone, were situated in different locations inside it. The irradiations were performed in uniform $^{60}$Co and $^{137}$Cs radiation fields, with the mobile phones connected and not connected to the cellular network. The dose decrease of 9% was measured for original resistors situated between layer of copper-clad laminate and battery, in comparison to the dose at the front of the phone. The resistors showed the lower signal when the mobile phone was connected to the cellular network, due to higher temperature inside the housing. The profile of fading was investigated within 3 month period for resistors irradiated with 1 Gy of gamma rays to estimate of the fading coefficient.

# INTRODUCTION

Widespread usage of ionizing radiation in health service and also in the industry stimulates the need for searching of fortuitous dosimeters, which can be used during unexpected exposition to radiation. This kind of dosimeters will be desirable for general public, which is not equipped with personal dosimeters. The possibility of rapid assessment of the radiation dose is very important for the prediction of health effects following an exposure.

Many scientists carried out researches on luminescence of everyday objects in the last few years. These objects should contain components sensitive to ionizing radiation and which could store the information about cumulated dose for several hours or days. For example, such items are: credit cards (Woda and Spottl, 2009, Mathur *et al.,* 2007), dental ceramics (Sholom *et al.,* 2011, Veronese *et al.,* 2010), different parts of electronic devices as resistors (Woda *et al.*, 2010, Bassinet *et al.,* 2010) and glass displays (Discher *et al.,* 2013, Discher and Woda, 2013, Piaskowska *et al.,* 2013). They could be considered as potential personal dosimeters, because they contain luminophors sensitive to ionizing radiation and usually are situated near the human body. The luminescence techniques like thermoluminescence (TL) and optically stimulated luminescence (OSL) can be applied to measure signals obtained from such materials. From all components of portable electronic devices the special attention can be paid to alumina ceramics components, such as resistors. These components are made in SMD (Surface Mounted Devices) technology and the substrates of resistors are composed of approximately 97% $Al_2O_3$ and 3% $SiO_2$ (Ekendahl and Judas, 2012). $Al_2O_3$:C detectors are very well known luminescent material applied in the OSL method (Boetter-Jensen *et al.,* 2003).

The aim of the work was to improve the precision of dose reconstruction, based on OSL properties of resistors from mobile phones. It was carried out by irradiation of several mobile phones of the same type, divided into three groups. First part of them were switched on and connected to the cellular network, second part - switched on but staying in the standby mode and the third part was switched off. The study of a dose distribution within the mobile phone was realized by measurements of additional resistors placed in different parts of the phone. The correction factor for fading of resistors OSL signal, measured during several days after exposition, was also calculated.

## MATERIALS AND METHODS

Investigations were carried out on chip resistors from NOKIA 3310, 6300, 6670 mobile phones. Before irradiation, additional commercially available resistors, similar in the shape and size to the resistors from circuit board of the studied mobile phones, were situated in different locations inside the mobile phones. After irradiation both types of resistors, the original as well as the additional ones, were extracted from the electronic circuit of the mobile phones. The resistors, having $Al_2O_3$ substrate, had dimensions of 1 x 0.5 x 0.35 mm. In each case one sample was composed of 10 resistors in order to increase the signal. The sample preparations were performed in the dark using red light.

Irradiations of mobile phones containing original and additional resistors were carried out with $^{90}Sr/^{90}Y$ source built-in in a Risoe reader, $^{60}Co$ from therapeutic Theratron 780E apparatus and $^{137}Cs$ sources installed in Laboratory for Calibration of Radiation Protection Instruments at the Institute of Nuclear Physics. The mobile phones were irradiated with doses in the range of 0.5 – 1.5 Gy.

The luminescence of resistors were investigated with the OSL method in an automated luminescent TL/OSL reader (model DA-20) produced by Risoe National Laboratory, Denmark. OSL measurements were done using blue diodes (470 nm ± 30 nm) with total power of 80 mW/cm$^2$ at the sample position. The readouts were performed with the U340 optical filter transmitting 250 - 400 nm light wavelength. The Risoe reader is equipped with a $^{90}Sr/^{90}Y$ source for calibration of samples. The OSL signal was measured at 100 °C following the pre-heat of 10 s at 120 °C.

## RESULTS AND DISCUSSION

### Mobile phone mode

One of the aim of investigation was related to the study how the mobile phone mode can impact a dose of ionizing radiation, which is absorbed by the resistors in electronic devices. The mobile phone mode means the phone status: if it is connected to the cellular network, remains in standby mode or is switched off. This experiment was carried out with 3 NOKIA 3310 mobile phones, which were irradiated in the uniform radiation field at the distance of 80 cm from a $^{60}Co$ (Theratron 780E). The conditions of exposure were the same for each mobile phone. As it was mentioned before, additional resistors were put into mobile phone under the battery.

Fig. 1 A and B present the shape of OSL curves for both type of resistors: original from a mobile phone (which were working during the connection) and additional ones, for different modes of mobile phone work. The signals are lower in the phones which were switched on for both type of resistors. However, this difference is higher in case of original resistors. Likewise, the measured OSL signals are weaker in case of additional resistors independent on the mobile phone mode, what is presented in the Fig. 1 A and B. The reason, why the signals are lower may be the absorption layer, which is the battery, and also the samples were measured at various time after irradiation. It means that the fading factor was not applied in this case.

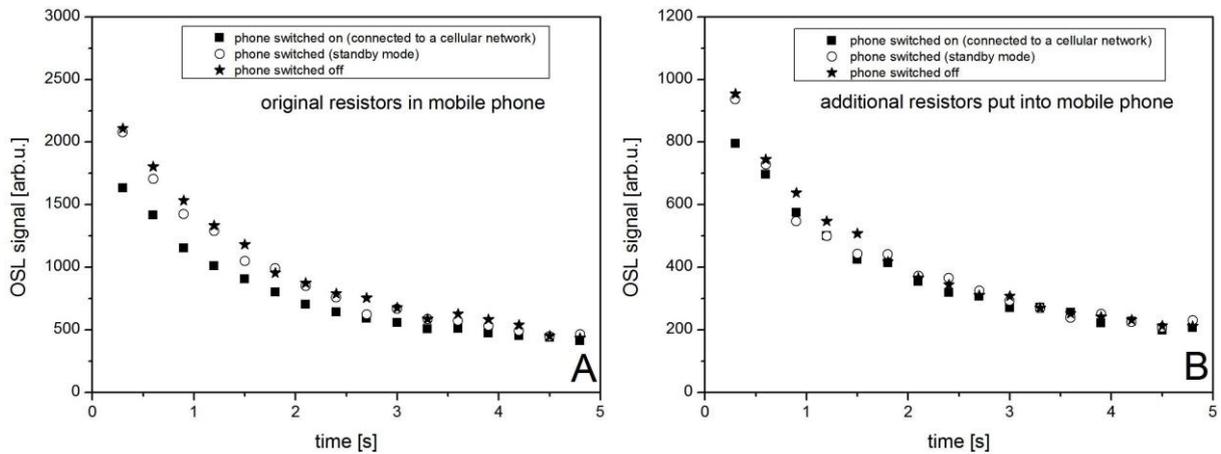

Fig. 1. OSL signal of original resistors (A) and additional resistors (B) for different types of mobile phone modes without fading correction.

**Fading factor**

The measured OSL signals were converted into the absorbed dose according to the relation:

$$Dose_{accidental} = \frac{OSL_{accidental}}{fading\ factor} \times calibration\ factor \quad (1)$$

Where the calibration factor is the ratio between the calibration dose and the signal absorbed in the detector during calibration exposure : $D_{calibration}/OSL_{calibration}$.

The signal loss during the time after irradiation is a significant disadvantage of the resistors. All of OSL data should be referenced to the time between irradiations and readouts by applying of the fading factor. The fading study was performed for nine groups of resistors consisting of five samples in each group. All samples were irradiated with $^{90}$Sr/$^{90}$Y source with the dose of 1 Gy and then OSL signal was measured at different times after irradiation up to 100 days (Fig. 2). The signal was normalized to the first measurement done immediately after irradiation. Every point corresponds to the mean value for a group of 5 resistors. The equation of fading is determined as follows:

$$OSL(time) = 1.11 - 0.07 \times \ln(time + 1.31) \qquad (2)$$

where time is expressed in minutes.

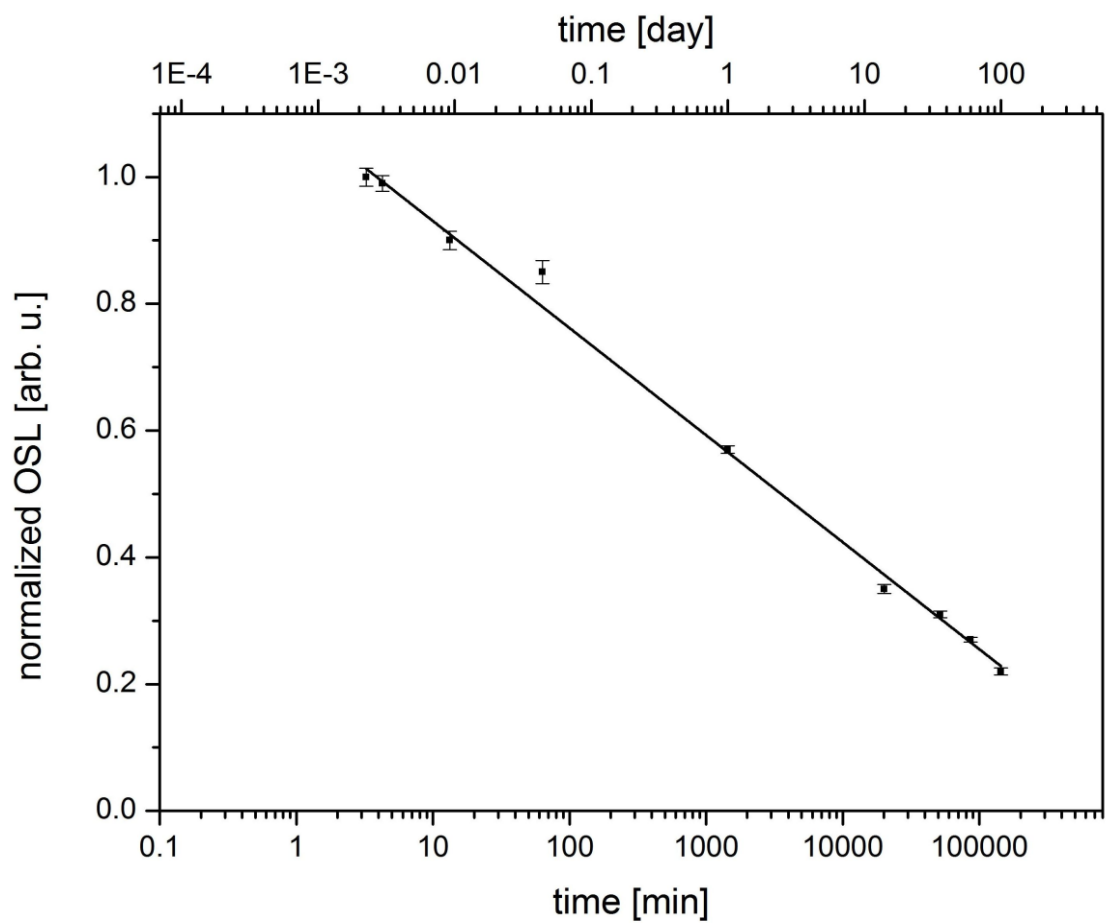

Fig. 2. Fading of OSL signal after exposition, normalized to first measurement. Fitted by Eqn (2).

After 3 months after irradiation the OSL signal amounts to 20% of the first recorded OSL signal and is still measurable.

The fading factor is expressed as:

$$f = \frac{OSL(time_{accidental})}{OSL(time_{calibration})} \quad (3)$$

where:

OSL($time_{accidental}$) – time between radiological accident and measurement;

OSL($time_{calibration}$) – time between calibration dose and measurement.

**Influence of the temperature**

Fading factor was applied to all calculations of absorbed doses. Fig. 3 shows that the absorbed dose in original and additional resistors is about 10% lower in case of the mobile phone connected to the cellular network during a whole irradiation time than in case of the mobile phone being in the standby mode or switched off.

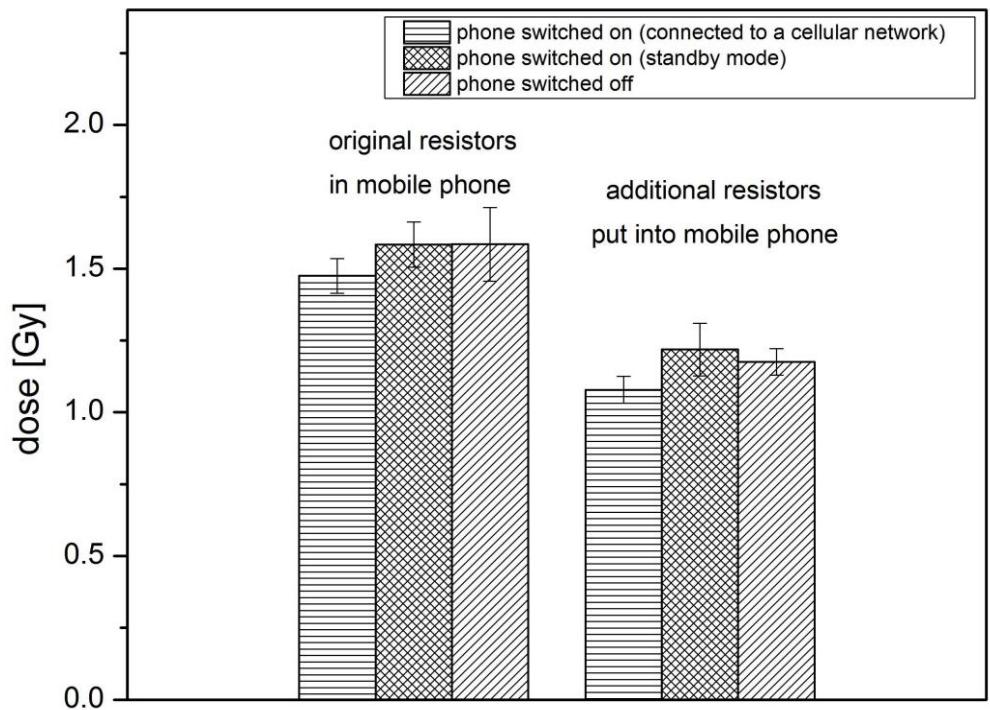

Fig. 3. The absorbed doses measured by original and additional resistors for different mobile phone modes.

It is known that when the mobile phone is in use during a conversation or listening to music, the temperature of mobile phone raises. It is difficult to estimate the temperature inside the mobile phone, but the temperature can be locally higher than outside and it can cause the reduction of dose. Because of this, another experiment was carried out on resistors. Additional, commercially available resistors were irradiated under the higher temperature conditions, using the $^{60}$Co source. Resistors were put into a special equipment – aluminum block (under the light protection), which were placed on an electric stove, perpendicular to the heating plate. The irradiations were performed after longer time needed for stabilization of the temperature. The temperatures of: 20 °C, 40 °C, 100 °C and 160 °C were applied. Fig. 4 shows correlation between decrease of the absorbed dose and temperature. Data showed that the dose absorbed by resistors is 9% lower, in case of the irradiation at 40 °C, than at room temperature and 90% lower for the irradiation at 160 °C. This may indicate that the temperature is responsible for reduced value of absorbed dose in resistors, in the case when the mobile phone is switched on and connected to the cellular network (Fig. 4). Also, it was possible to find that the absorbed dose in the additional resistors is still lower than in the original even after fading correction. The difference could disappear after application of the factor for different type of layers.

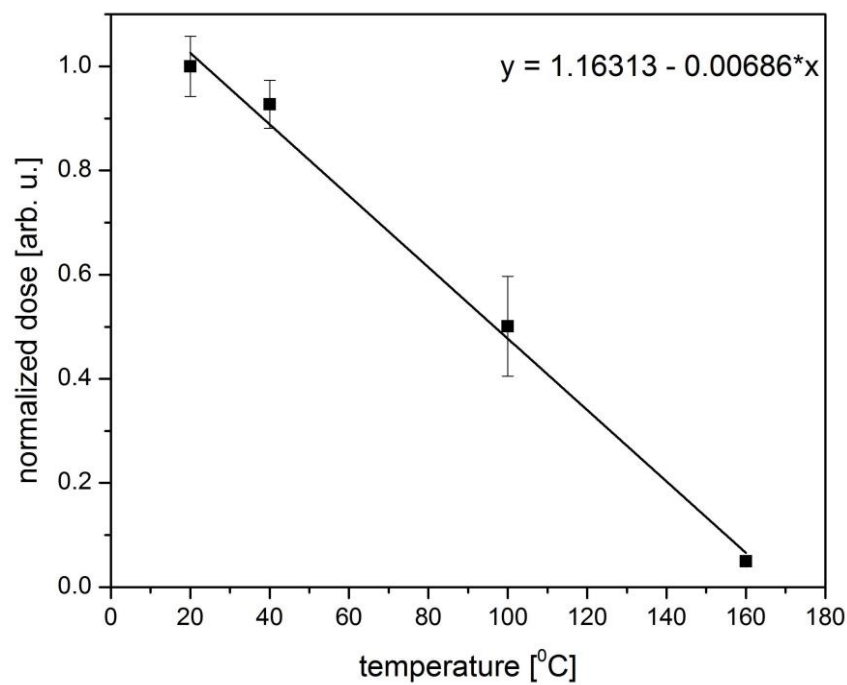

Fig. 4. Absorbed dose versus irradiation temperature. Experiment was carried out with additional resistors put into alumina blocks placed on electric stove.

**Dose distribution inside of a mobile phone**

The second part of the investigation was focused on the determination how the readout dose depends on the position of the resistor inside of the mobile phone housing. It was realized by OSL measurements of additional resistors placed in different locations inside of the mobile phone, while the original resistors are usually located between the layer of copper-clad laminate and battery. The additional resistors were put into mobile phone in five different locations, i.e. in front of the special layers: in front of the mobile phone, glass display, layer of the copper-clad laminate – at the level of original resistors, in front and behind of the battery. In each layer the number of resistors was 40. This experiment was carried out using two mobile phones: NOKIA 6300 and NOKIA 6670, which were irradiated with $^{137}$Cs source. The average thickness of mobile phones was 1.7 cm. The individual layers of the resistors were placed in 0.4, 0.8, 1.3, 1.7 cm from the mobile phone front (glass display), respectively. The dose measured by the resistors is about 23% lower at the distance of 1.7 cm in comparison to the dose absorbed at the front of the mobile phone (Fig. 5).

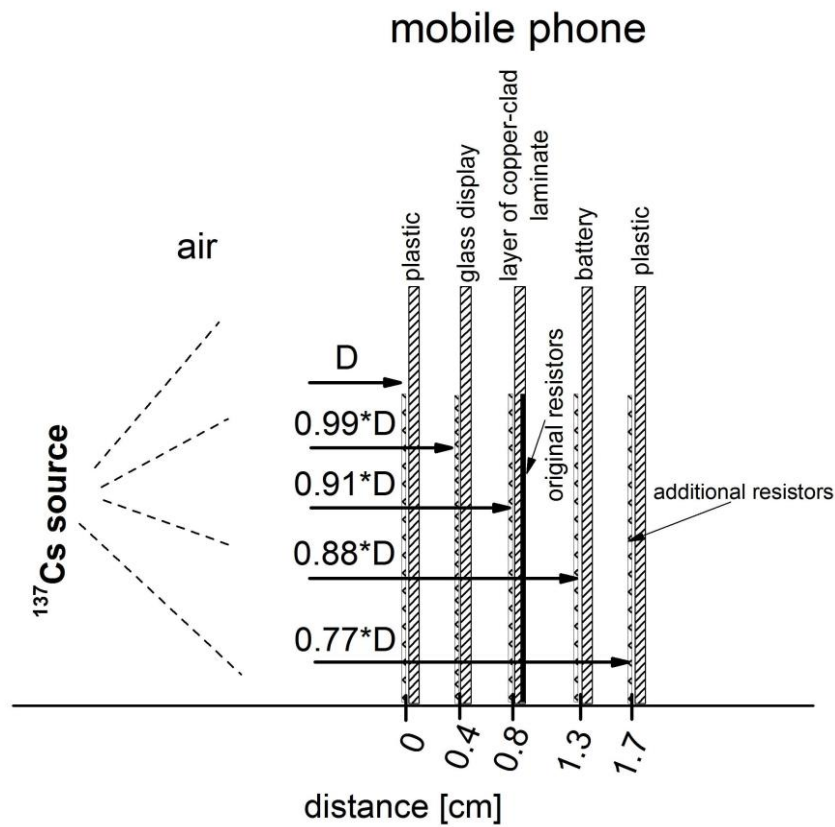

Fig. 5. Dose absorbed by resistors placed at different layers in a mobile phone (from mobile phone front to back housing).

This is due to absorption of radiation by the different layers of the mobile phone. The similar situation occurs for the dose absorbed in the original resistors. The absorbed dose in original resistors is about 9% lower than that in the resistors placed in front of the mobile phone. It means that the readout of original resistors coming from the mobile phone interior is at least of about 9% less than the signal at the front of the mobile phone and should be corrected regarding to the resistor position.

**CONCLUSIONS**

Investigation of OSL signal of resistors confirmed that such devices like mobile phones could be considered as the useful tools for dose estimation after unexpected radiological accident. The mobile phones, possessing own original resistors and additional resistors located at different places, were tested. The mobile phones kept in three modes of work: connected to the cellular network, remaining in a standby mode or switched off were irradiated. The investigation showed that the dose absorbed in original and additional resistors is lower in case of the mobile phone being in use than switched off. It is probably caused by a higher temperature of particular parts inside of the mobile phone during its usage. Determination of dose homogeneity in a whole volume of the mobile phone was measured by additional resistors placed at different phone layers. The dose absorbed in original resistors was about 9% lower than in front of the mobile phone.

The better precision of the assessment of the doses reconstructed from mobile phone resistors can be achieved by allowance of the factors for phone work mode, resistor position and fading.

**ACKNOWLEDGEMENTS**

The work was supported by the National Centre for Research and Development (Contract No. PBS1/A9/4/2012). Anna Mrozik has been partly supported by the EU Human Capital Operation Program, Polish Project No. POKL.04.0101-00-434/08-00.